 \documentclass{aa}

\usepackage{threeparttable}
\usepackage{graphicx}
\usepackage{amsmath}
\usepackage{xspace}
\usepackage{subcaption} 
\usepackage{upgreek}
\usepackage{float}

\usepackage{tablefootnote}
\usepackage{tikz}

\usepackage{txfonts}

\newcommand\nustar{\textit{NuSTAR}\xspace}
\newcommand\nustarf{\textit{Nuclear Spectroscopic Telescope Array}\xspace}
\newcommand\suzaku{\textit{Suzaku}\xspace}

\newcommand\swiftf{\textit{Neil Gehrels Swift}\xspace}
\newcommand\swift{\textit{Swift}\xspace}

\newcommand{\cen}{\mbox{Cen~X-3}\xspace}
\newcommand{\her}{\mbox{Her~X-1}\xspace}

\bibpunct{(}{)}{;}{a}{}{,} 

\defcitealias{BW07}{BW07}

\begin{document} 

\title{Fitting strategies of accretion column models and application to the broadband spectrum of \cen}

\authorrunning{P.\ Thalhammer et al.}
\titlerunning{Fitting strategies of the BW07 model}

   \author{Philipp Thalhammer
          \inst{1}
          \and
          Matthias Bissinger\inst{1,2}
          \and
          Ralf Ballhausen\inst{1,12}
          \and
          Katja Pottschmidt\inst{3,4}
          \and
          Michael T.\ Wolff\inst{5}
          \and
          Jakob Stierhof\inst{1}
          \and
          Ekaterina Sokolova-Lapa\inst{1,11}
          \and
          Felix F\"urst\inst{6}
          \and
          Christian Malacaria\inst{7,8}
          \and
          Amy Gottlieb\inst{9}
          \and 
          Diana~M.~Marcu-Cheatham \inst{3,4}
          \and
          Peter A. Becker \inst{10}
          \and 
          J\"orn Wilms\inst{1}
        }

        \institute{Dr.\ Karl Remeis-Observatory and Erlangen Centre
          for Astroparticle Physics,
          Universit\"at Erlangen-N\"urnberg, Sternwartstr.~7, 96049
          Bamberg, Germany
          \and
          Erlangen Centre for Astroparticle Physics, Universit\"at
          Erlangen-N\"urnberg, Erwin-Rommel-Stra\"se 1, 91058
          Erlangen, Germany
          \and
         Department of Physics and Center for Space Science and Technology,
         University of Maryland Baltimore County, 1000 Hilltop Circle, Baltimore, MD 21250, USA
          \and
          CRESST and NASA Goddard Space Flight Center, Astrophysics
          Science Division, Code 661, Greenbelt, MD 20771, USA
          \and
          Space Science Division, Naval Research Laboratory,
          Washington, DC 20375-5352, USA
          \and
          Quasar Science Resources SL for ESA, European Space Astronomy Centre (ESAC), Science Operations Departement, 28692 Villanueva de la Ca\~nada, Madrid, Spain 
          \and
          NASA Marshall Space Flight Center, N SSTC, 320 Sparkman Drive,
          Huntsville, AL 35805, USA \and Universities Space Research
          Association, NSSTC, 320 Sparkman Drive, Huntsville, AL 35805,
          USA
        \and Department of Astronomy, University of Florida, Gainesville, FL 32611, USA
        \and Physics and Astronomy, George Mason University, Fairfax, VA,
        \and Sternberg Astronomical Institute, M.~V.~Lomonosov Moscow State University,
        Universitetskij pr., 13, Moscow 119992, Russia
        \and Southeastern Universities Research Association, Washington, DC 20005 USA 
        }

        \date{RECEIVED: ACCEPTED:}

  \abstract
  { Due to the complexity of modeling the radiative transfer inside
    the accretion columns of neutron star binaries, their X-ray
    spectra are still commonly described with phenomenological models, for example, a cutoff power law. While the behavior of these models is
    well understood and they allow for a comparison of different
    sources and studying source behavior, the extent to which the
    underlying physics can be derived from the model parameters is
    very limited.  During recent years, several physically motivated
    spectral models have been developed to overcome these limitations.
    Their application, however, is generally computationally much more
    expensive and they require a high number of parameters which are
    difficult to constrain. Previous works have
    presented an analytical solution to the radiative transfer
    equation inside the accretion column assuming a velocity profile
    that is linear in the optical depth. An implementation of this solution that is both fast and accurate enough to be fitted to
    observed spectra is available as a model in XSPEC. The main difficulty of
    this implementation is that some solutions violate energy
    conservation and therefore have to be rejected by the user.  We
    propose a novel fitting strategy that ensures energy conservation
    during the $\chi^2$-minimization which simplifies the application
    of the model considerably. We demonstrate this approach as well as
    a study of possible parameter degeneracies with a comprehensive
    Markov-chain Monte Carlo analysis of the complete parameter space
    for a combined \nustar{} and \swift{}/XRT dataset of Cen X-3. The
    derived accretion-flow structure features a small column radius of
    $\sim$63\,m and a spectrum dominated by bulk-Comptonization of
    bremsstrahlung seed photons, in agreement with previous studies.
  }

   \keywords{methods: data analysis – X-rays: binaries – stars: individual (Cen X-3) – stars: neutron }

\maketitle
%
\section{Introduction}

X-ray pulsars are powered by the release of gravitational energy of material
falling from an optical companion onto the surface of an accreting neutron star (NS).
Due to their strong magnetic field, the infalling material is funneled onto the
magnetic poles where it is decelerated and its kinetic energy is released in the
form of radiation \citep[see][for a general overview of the observational
properties of such systems]{basko_1976, wolff_19,staubert2019a}. The physical
mechanism decelerating the infalling material depends on the mass accretion
rate, $\dot{M}$, and thus the luminosity \citep[e.g.,][and references
therein]{davidson73,becker_2012,mushtukov:2015b}. Above a certain critical
luminosity the matter is decelerated predominantly by radiative pressure, which
leads to the formation of a radiative shock above the polar caps of the NS
\citep[e.g.,][and references therein]{becker_2012,mushtukov:2015b}, while at
lower luminosities, as seen in X-Per, for example, other means of deceleration such as
a classical collision-less shock have been brought forward
\citep{Shapiro_75,langer_82}. In this environment photons gain energy through
thermal as well as bulk motion Comptonization while propagating through the
accretion column. The dominant sources of seed photons for this process are
bremsstrahlung throughout the column, black-body radiation from the heated
accretion mound, and radiative de-excitation of collisionally excited Landau
levels \citep[][hereafter BW07]{BW07}.

One source particularly suited for the study of the the spectral formation of
these objects is \cen, due to its bright and rather persistent nature
\citep[although with a known strong long-term flux variation, see,
e.g.,][]{paul_2005} and a prominent cyclotron resonant scattering
feature (CRSF) at ~30 keV \citep{nagase_1992,santangelo_98}. \cen is an
accreting high mass X-ray binary (HMXB) with a spin period of
${\sim}4.8$\,s and an orbital period of 2.1\,d \citep{cen_72}.
Discovered in 1971 with the \textit{Uhuru} satellite \citep{cen_71},
\cen was the first X-ray source identified as an X-ray pulsar. The
binary system is located at a distance of $5.7\pm1.5\,\mathrm{kpc}$
\citep{distance57} and consists of a NS with a mass of
$1.2 \pm 0.2\,\mathrm{M}_\odot$ and O6--8 III supergiant star with a
mass of $20.5 \pm 0.7\,\mathrm{M}_\odot$ \citep{compan_99}. The system
is at high inclination, such that the NS is eclipsed by its
donor star for around 20\% of the orbital period \citep{suchy_08}. The
primary mode of mass transfer in the system is Roche-lobe overflow,
leading to the formation of an accretion disk around the NS
\citep{Tjemkes86}. Its X-ray spectrum has been well described by a
cut off power-law with $\Gamma\sim 1$; strong full and partial
absorption due to the presence of dense clumps of material originating
from the stellar wind and passing through the line of sight to the
NS \citep{suchy_08}; three fluorescence lines, which
previous studies have traced back to regions of varying degrees of
ionization and distance to the NS
\citep{ebisame_asca_1996,suchy_08}; a ``10-keV'' feature, of which the
origin is still unclear \citep{suchy_08,marcu_2020}; and the CRSF around
30\,keV \citep{santangelo_98}.

The general problem of spectrum formation in the accretion column of
sources such as \cen is highly complex. It requires modeling of the
radiative transfer and plasma deceleration, moderated by the radiation
field, and can only be addressed numerically
\citep[see, e.g.,][]{wang_1981,west_2017,Gornostaev_2021,mushtukov:2015b,Kawashima_2020}. The associated computing
times are prohibitively large, such that a direct comparison with
observations is not feasible. Consequently, the observational
literature has mainly concentrated on the empirical description of the
spectral properties and variability of accreting X-ray binaries
\citep[see, e.g.,][and references therein]{mueller_0115}, and the
number of attempts to connect observations with physical parameters of
the accretion column is still small \citep[see, e.g.,][for some
pionieering examples]{Ferringo_09,wolff_16}.

\citet{BW07} and \citet{Farinelli_2012} showed, however, that under
certain simplifying assumptions regarding the velocity profile of the
infalling matter, and the details of photon-electron interactions, an analytical solution of the radiative transfer
problem in the accretion column can be obtained.
This model provides a spectrum emitted from the
 accretion column, however omitting the possible reflection of the
 photons from the NS surface and the resonant Compton
 scattering that is responsible for the formation of cyclotron lines.
Implementations of
the analytical solution by BW07 for common X-ray analysis tools such
as \texttt{ISIS}
\footnote{\url{https://space.mit.edu/CXC/isis/}}\citep{houck_2000} or
\texttt{XSPEC}\footnote{\url{https://heasarc.gsfc.nasa.gov/xanadu/xspec/}}
\citep{arnaud_1996} are publicly available.
\footnote{\url{https://www.isdc.unige.ch/~ferrigno/images/Documents/BW_distribution/BW_cookbook.html}}
The specific implementations \texttt{BWsim} (used throughout this
work) and \texttt{BWphys} represent the same physical model, but with
different parameterizations. By nature of the complex physical
processes addressed by the BW07 model, it often has a larger number of
parameters than comparable phenomenological models, some of which are
additionally affected by degeneracies. The application to
observational data is therefore still very challenging and in this
regard similar to other physically motivated spectral models such as,
for example, advanced plasma models applied to high-resolution X-ray
spectra. A particular practical difficulty in the usage of the BW07
model is that while single parameter ranges can be estimated quite
reliably, individual parameter combinations may violate underlying
model assumptions. The example which we are focusing on is an accretion
rate and the related ``accretion luminosity''. The latter can differ
significantly from the luminosity estimate based on the bolometric
flux.

In this paper we discuss fitting strategies of the BW07 model
addressing this difficulty and demonstrate an application to
contemporaneous \swiftf(\swift) and \nustarf(\nustar) observations of
\cen.
We suggest that the developed fitting approach can also be used
for  other models that require complex constraints on the
parameter landscape. The remainder of the paper is organized as follows. In
Sect.~\ref{obs_red} we briefly discuss the used \textit{Swift} and
\textit{NuSTAR} observations of \cen and the respective data
reduction. We then discuss in Sect.~\ref{meth} the technical
challenges of applying the BW07 model to observational data and how to
address these with a new spectral fitting approach. We present the
result of this application to the broadband spectrum \cen in
Sect.~\ref{res}. A discussion and conclusions of the results are
presented in Sect.~\ref{disc} and Sect.~\ref{conc}, respectively.

\section{Observation \& data reduction} \label{obs_red}

\begin{figure}
  \resizebox{\hsize}{!}{\includegraphics{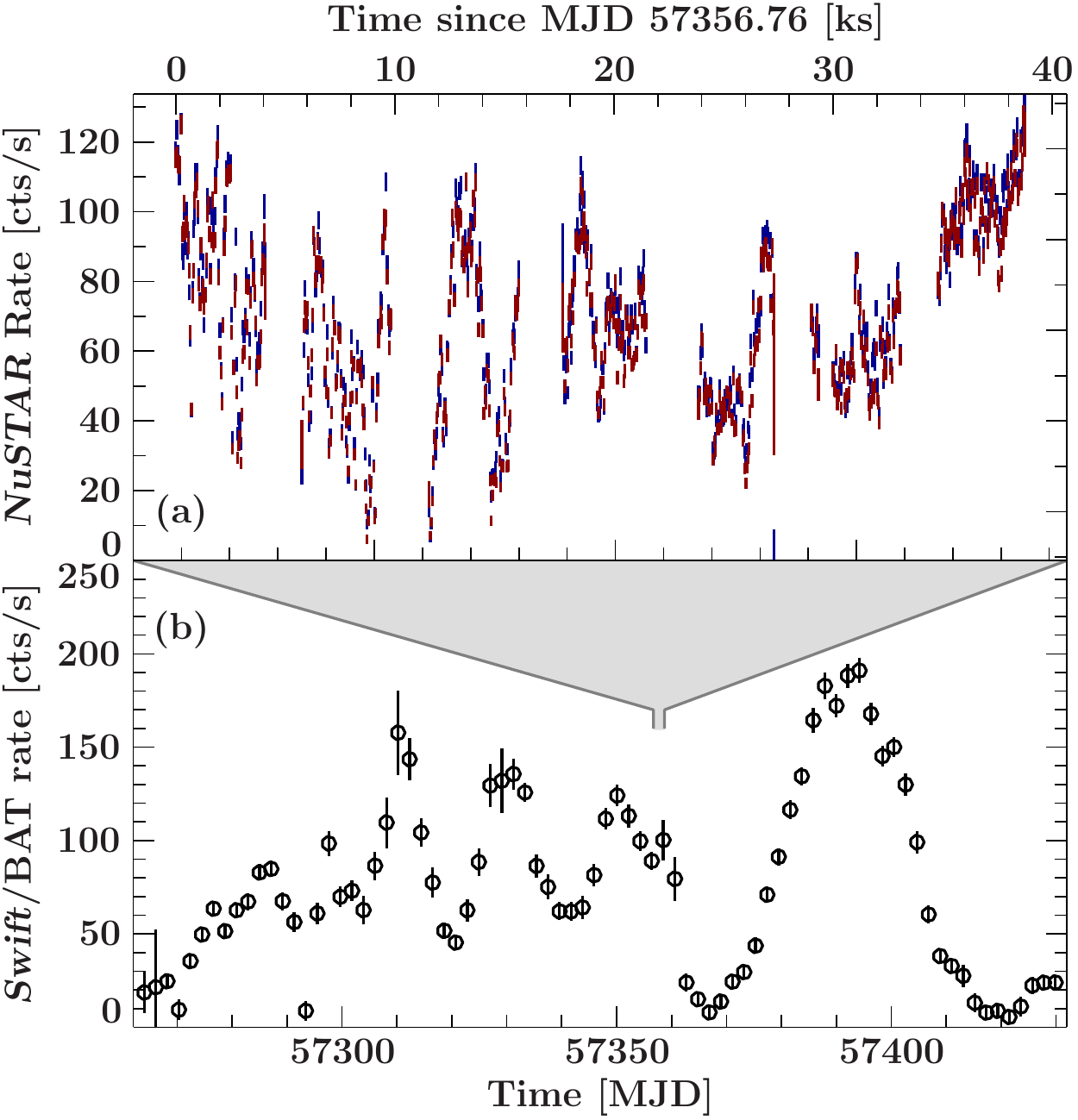}}
  \caption{\textit{(a)} Light curve of the observation as seen by \nustar. Red and blue data points show the \nustar light-curve corresponding 
to  FPMA and FPMB.
    \textit{(b)} \swift/BAT \citep{krimm_13} light-curve  of
    \cen around the observation time, binned to the
    orbital period of ${\sim}2.08\,\mathrm{d}$. }\label{fig:lc}
\end{figure}

On 2015 November 30, \textit{NuSTAR} \citep{NuStar} observed \cen with
a total exposure time of 21.4\,ks and 21.6\,ks for the two focal plane
models, FPMA and FPMB, respectively (ObsID 30101055002). The
\swift/BAT light curve at the time of observations is shown in
Fig.~\ref{fig:lc}. We extracted these data with the official
\textit{NuSTAR} analysis software contained in HEASOFT 6.22 and CALDB v20180419, using
source regions with a $120''$ radius around the source position.
Background spectra were accumulated using regions with the same radius
in the southern part of \textit{NuSTAR}'s field of view.

In order to investigate the spectrum down to 1\,keV we include
contemporaneous \swift data \citep{Swift} taken taken on 2015 December
10 (ObsID~00081666001) with a total exposure time of 425\,s in Photon
Counting Mode (PC) and 1522\,s in Windowed Timing mode. To mitigate
the significant pile up in the PC mode data, we defined the source
region as an annulus with inner radius of 8\,pixel and outer radius of
30\,pixel, and use a 60\,pixel circle offset from the source to
extract the background. In Windowed Timing mode \textit{Swift}-XRT has
only one dimensional imaging capabilities. For this mode, the central
20 pixels were used as the source region. A region between 80 pixel
and 120 pixel away from the source on both sides was used as the
background.

In the following spectral analysis, which was performed with
\texttt{ISIS} version 1.6.2-41 \citep{houck_2000}, we use the
\textit{NuSTAR} data in the energy band between 3.5\,keV and 79\,keV,
while the \textit{Swift} data were considered in the band from
1.0\,keV to 10\,keV.

\section{Fitting the accretion rate of \texttt{BWsim}} \label{meth}
\subsection{The accretion column model by \citet{BW07} }
A detailed description of the accretion model \texttt{BWsim} is given by BW07
and \citet{wolff_16}, so we only summarize its key features: The gas
entering the cylindrical column from the top is decelerated in an extended
radiation-dominated standing shock region and eventually settles on the thermal
mound at the bottom of which it reaches a velocity of zero. In this column, the
deceleration of the accreted matter leads to the emission of photons.
Specifically the following three types of seeed photons are considered by BW07:
Blackbody radiation from the hot thermal mound\footnote{We note that the
     top of the thermal mound is not the same as the NS surface.
     Therefore matter enters the mound with a certain inflow speed heating the
     mound to a temperature given by Eq. 93 in BW07.} at the bottom of the
     accretion column;
  Bremsstrahlung emitted by the electrons streaming along the magnetic field
     (free-free emission); and
     Cyclotron radiation produced by the decay of electrons collisionally
   excited to the first Landau level\footnote{These seed photons are treated by
   the model as simply monochromatic without  intrinsic line broadening. The
   strong broadening due to thermal and bulk Componization that the seed photons
   experience subsequently after their injection would likely wash out the
   intrinsic broadening of the cyclotron seed photon. Hence the neglect of the
   intrinsic broadening is probably reasonable. }.

As they propagate through the column, these photons then interact with
the accreted material. For the radiative transport one needs to
account for both bulk and thermal Comptonization of seed photons
(i.e., first- and second-order Fermi energization due to collisions
with gas deceleration along the $z$-axis), which leads to a
two-dimensional radiative transfer equation. The photon distribution
$f(z,\epsilon)$ is a function of the height $z$ and the photon energy
$\epsilon$ and, as given by BW07 (Eq.~15), the transport equation is
given by
\begin{multline}
\frac{\partial f}{\partial t} + v \, \frac{\partial f}{\partial z}
= \frac{dv}{dz}\,\frac{\epsilon}{3} \,
\frac{\partial f}{\partial\epsilon}
+ \frac{\partial}{\partial z}
\left(\frac{c}{3 n_\mathrm{e} \sigma_\parallel}\,\frac{\partial f}{\partial z}\right)\\
- \frac{f}{t_{\rm esc}}
+ \frac{n_\mathrm{e} \Bar{\sigma} c}{m_\mathrm{e} c^2} \frac{1}{\epsilon^2}
\frac{\partial}{\partial\epsilon}\left[\epsilon^4\left(f
+ k T_\mathrm{e} \, \frac{\partial f}{\partial\epsilon}\right)\right]
+ \frac{Q(z,\epsilon)}{\pi r_0^2}
\ ,
\label{eq3.1}
\end{multline}
where $z$ is the distance from the stellar surface along the column
axis, $v$ is the inflow velocity, $t_\mathrm{esc}$ represents the mean
time photons spend in the plasma before diffusing through the walls of
the column, $ \sigma_\parallel$ is the electron scattering
cross-section parallel to the magnetic field, $\Bar{\sigma}$ is the
angle-averaged mean scattering cross-section and $Q(z,\epsilon)$
denotes the photon source distribution.

The left-hand side of Eq.~\eqref{eq3.1} describes the time derivative
of the photon density in the co-moving reference frame, while the
right-hand side accounts for (from left to right) bulk Comptonization
inside the radiative shock region, diffusion of photons along vertical
column axis, escape of photons though the column walls, thermal
Comptonization described by the Kompaneets operator
\citep{kompaneets_57}, and the injection of seed photons into the
accretion column.

The escape time $t_\mathrm{esc}$, which quantifies the diffusion of photons
through the column walls, is only a function of height along the column and is,
similar to the scattering cross-sections, already averaged over polarization
states and photon energies. However, as indicated by Eq. 18 from BW07, the
escape time utilized the perpendicular scattering cross section $ \sigma_\perp$,
and is therefore sensitive to the direction of propagation. To make an
analytical solution feasible it is necessary to assume a temperature profile
that is constant over the accretion column. There have been numerical
calculations showing significant temperature variation inside the column
\citep{mushtukov:2015b,west_2017}. However, one would expect that
inverse-Compton “thermostat” to keep the electron temperature roughly constant
in the region of the accretion column where Comptonization is strongest, which
is the same region that generates most of the observed radiation.  

One should also note that resonant scattering is not included in the
BW07 model. This is probably reasonable for sources with relatively
strong magnetic fields, so that the cyclotron energy is around
$\sim$40 keV or higher, because in this case most of the photon
scattering occurs away from the resonance, in the continuum region of
the cyclotron cross section (see discussion just before Eq.~6 in
BW07).

The escaping spectrum, which is derived from the photon distribution $f$ and
$t_\mathrm{rsc}$ using Eq. 69 of BW07, is integrated along the column
height to yield the detectable spectrum.

BW07 also showed that the assumption of a velocity profile linear in
optical depth, $\tau$, results in a radiation transfer equation that
is separable in energy and space, which in turn allows the derivation
of a Green's function which can then be applied to the three seed
photon distributions mentioned above. While the resulting computing
times are still large compared to phenomenological models, they are
small enough that direct comparisons between the model and data are
possible. In the resulting model the free parameters are the mass
accretion rate, $\dot{M}$, the column radius, $r_0$, the magnetic
field strength responsible for the cyclotron seed photons, $B_0$, the
electron temperature of the Comptonizing electrons $T_\mathrm{e}$, and
two so-called similarity parameters\footnote{The alternative
  parameterization of the model, \texttt{BWphys}, uses the parallel and
  average scattering cross sections, $\sigma_\parallel$ and
  $\Bar{\sigma}$, instead of the similarity parameters $\xi$ and
  $\delta$.} $\delta$ and $\xi$, which describe properties of the
accretion flow and of the radiative transfer in the accretion column
\citep[see Eq.~3 and 4 of][]{wolff_16}. Specifically, $\xi$ is related
to the ratio of the accretion and photon escape timescales below the
shock
\begin{equation}
  \xi=4.2\frac{t_\mathrm{shock}}{t_\mathrm{escape}}\ =\frac{\pi r_{0} m_\mathrm{p} c}{\dot{M} \sqrt{\sigma_{\perp} \sigma_{\|}}},
\end{equation}
where $m_\mathrm{p}$ is the proton mass, and $\delta$ is related to
the Compton-$y$-parameters for bulk motion and thermal Comptonization,
\begin{equation}
  \delta=4\frac{y_\mathrm{bulk}}{y_\mathrm{th}}=\frac{\alpha}{3} \frac{\sigma_{\|}}{\bar{\sigma}} \frac{m_{e} c^{2}}{k T_{e}} ,
\end{equation}
where
\begin{equation}
  \alpha=\frac{32 \sqrt{3}}{49 \ln (7 / 3)} \frac{G M_{*} \xi}{R_{*} c^{2}},
\end{equation}
and where $ M_{*}$ and $R_{*}$ are the mass and radius of the NS,
$\sigma_{\|}$ is the average scattering cross-section perpendicular to
the magnetic field, and $y_\mathrm{bulk}$ and $y_\mathrm{th}$ are the
Compton $y$-parameters as defined by \citet{Rybicki_79}. In order to
solve Eq.~\eqref{eq3.1} analytically, in addition to the approximate
velocity profile, BW07 assume a uniform temperature cylindrical
accretion flow in a constant $B$-field, which is in a steady state and
consists of a fully ionized hydrogen plasma. The electron
cross-sections are approximate and estimated for electrons propagating
parallel or perpendicular to the magnetic field and interacting with
photons having the mean photon energy, that is averaged over the photon
energy. These assumptions limit the application to strongly magnetized
NS at high accretion rates, where radiation pressure plays a
dominant role in the deceleration. This should make the model
applicable down to luminosities around
$L_\mathrm{x} \sim 10^{35\ldots 36} \mathrm{erg}\,\mathrm{s}^{-1}$,
but an assessment of the validity of the model assumptions is
recommended on a case-by-case basis.
An overview of the resulting free model parameters is shown in Table \ref{bw07_par}.
\begin{table}
\centering
\begin{threeparttable}

\caption{Overview of the free parameters of the accretion column model by \citet{BW07}}   \label{bw07_par}
 \begin{tabular}{lll}
  Parameter & Unit & Description  \\
    \hline
   $ \dot{M}$ & [g\,s$^{-1}$] &    Accretion rate    \\
   $ kT_\mathrm{e} $ & keV & Mean electron temperature   \\
   $ r_0 $ & m & Accretion column radius    \\
   $ B $&  G & Magnetic field along the column   \\
   $ \xi ^\dagger$ &   & Importance of the escape of  \\
                    & & photons from the accretion column  \\
    $ \delta ^\dagger$ &    & Relative importance of bulk    \\
                        & &  \& thermal  Comptonization \\
   \end{tabular}
   \begin{tablenotes}[flushleft]
     \item \tiny Note: $^\dagger$ Equivalent to a parametrization using $\bar{\sigma}$ \& $\sigma_\parallel$
   \end{tablenotes}
    \end{threeparttable}
  \end{table}

\subsection{Iterative approach for energy conservation}
A fundamental difficulty of the application of the BW07 model
\citep[which is also present in the model of][]{Farinelli_2012} when
fitting X-ray spectra is that for many parameter combinations it lacks
self-consistency and does not automatically conserve energy.
Specifically, there can be a mismatch between the integrated ``X-ray
luminosity'',
\begin{equation}\label{eq:lx}
  L_\mathrm{X} = 4 \pi\, D^2 \int_{E_\mathrm{min}}^{E_\mathrm{max}} S_E(E)\, dE
\end{equation}
 and the accretion luminosity
\begin{equation}\label{eq:acclum}
  L_\mathrm{acc} = \frac{\dot{M} M_{*} G}{R_{*}}
\end{equation}
as estimated from the loss of potential energy of the accreted
material, assuming that all accretion energy is radiated away. In
Eqs.~\eqref{eq:lx} and \eqref{eq:acclum} $S_E(E)$ is the
model-predicted photon flux adjusted to the rest frame of the
accretion column\footnote{In order to calculate the flux close to the
  NS surface, where the spectrum emerges, the red-shift of
  $z$=0.3, which is otherwise applied, is set to 0, as well as the
  column density of both the partial and complete absorption model.},
$E_\mathrm{min}$ and $E_\mathrm{max}$ are appropriate energy bounds
for the integration over the energy $E$, $D$ is the distance to the
source, $\dot{M}$ is the mass accretion rate, $M_{*}$ and $R_{*}$ are
the NS's mass and radius, respectively, and $G$ is the
gravitational constant. We note that the $\dot{M}$ is not just a
scaling parameter, but it is also connected to the shape of the
emitted spectrum, since, e.g., the cyclotron and bremsstrahlung seed
photon generation is a function of density and hence $\dot{M}$,
whereas the blackbody seed photon injection only depends on the size
and temperature of the polar cap. The mass accretion rate parameter is
therefore correlated to some extent with other model parameters
determining the spectral shape and cannot be constrained uniquely from
the observed flux. Based on the argument of energy conservation,
parameter combinations resulting in good description of the spectral
shape but with a strong mismatch of the accretion and X-ray luminosity
therefore have to be rejected. Since these invalid solutions cannot be
avoided a priori, e.g., by choosing appropriate 1D parameter
boundaries, other strategies of ensuring consistent parameter
solutions during the fitting procedure are required. While such higher
dimensional constraints on the parameter space can be accounted for
quite directly in fitting algorithms that sample the parameter space
(e.g., Markov Chain Monte Carlo) by rejecting invalid parameter
combinations, typical $\chi^2$-minimization algorithms require a
certain integrity of the parameter space. \citet{wolff_16} therefore
propose an interactive approach to ensure energy conservation manually
when comparing the BW07 model to observations. This approach is
typically done by estimating $\dot{M}$ from the source spectrum by
positing that $ L_\mathrm{X}$ equals $L_\mathrm{acc}$. Holding this
$\dot{M}$ fixed, the model is then fit to the data, obtaining a new
set of best-fit parameters. This fit, however, will generally change
the integrated luminosity of the model. The new fit is therefore not
consistent with the initial condition of
$L_\mathrm{X}=L_\mathrm{acc}$. One then iterates the fit by adjusting
the parameters until equality of both parameters is reached.

In practical terms, to determine a first guess for the accretion rate,
the observed spectrum is initially fitted with an empirical model,
i.e., some type of power-law with an exponential cut-off \citep[see,
e.g.,][for a description of the relevant models]{mueller_0115}. One
then uses this fit and the known distance to calculate an initial $L_\mathrm{X}$,
and from Eq.~\eqref{eq:lx} and~\eqref{eq:acclum} then obtains an
an initial accretion rate $\dot{M}$\footnote{This assumes the released
energy is emitted isotropically. We address this assumption in section \ref{disc}.}.
This estimate is then followed by a first
$\chi^2$-minimization using a \texttt{BWsim} continuum with $\dot{M}$
held fixed at  this value. The resulting flux is then used
to derive a new value of $\dot{M}$, and this approach is iterated
until convergence is reached in $\dot{M}$, i.e., once the relative
change in $\dot{M}$ becomes smaller than a certain threshold, here
1\%. For the joint \textit{Swift}/\textit{NuSTAR} data of \cen this
method converges after 3--6 iterations.
In practice, this method is a very cumbersome approach that requires
significant manual intervention and ``baby sitting'' of the fits,
hindering the application to larger samples of spectral data. Another
issue with this iterative approach is that during each minimization
step $\dot{M}$ is treated as a fixed parameter and is, therefore, not
allowed to deviate from the value derived from the model luminosity.
This approach neglects the fact that, e.g., the distance to the source
is not precisely known, which would result in a systematic shift in
the assumed accretion rate.

\subsection{Biasing the fit statistics}

\begin{figure}
  \resizebox{\hsize}{!}{\includegraphics{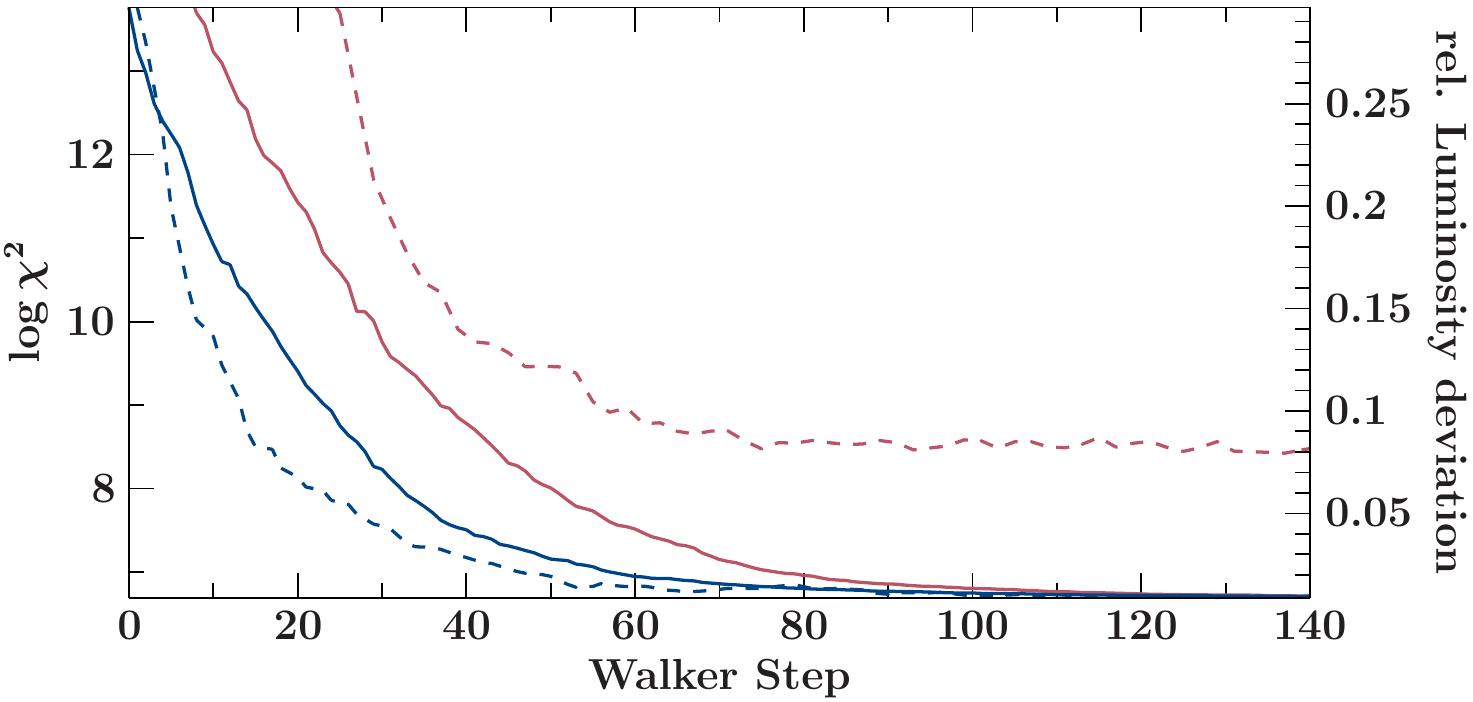}}
  \caption{Convergence behavior during the MCMC run with (blue) and without
  (red) the constraint. The dotted line shows the median  of the relative luminosity deviation ( $ \left| L_\mathrm{X} -  L_\mathrm{acc} \right| / L_\mathrm{X} $ ) for the walkers of each MCMC step. The
  solid line indicates, for both runs, the convergence of the regular, unmodified $\chi^2$
  median of each step. We note that the convergence of the unmodified fit-statistic
  does not deteriorate while only with our constraint the criterion of energy
  conservation is quickly reached. }\label{fig:con}
\end{figure}

To avoid the interactive procedure we propose an alternative strategy to address
this issue which does not require an iterative approach but rather introduces
energy conservation directly into the data modeling. The idea of our technique
is to let $\dot{M}$ vary during the fit, but to bias the fit statistics in a way
that it disfavors values of $\dot{M}$ which do not fulfill energy conservation,
i.e., for which $L_\mathrm{X} \neq L_\mathrm{acc}$. In \texttt{ISIS}, this
approach can easily be implemented as one has a direct programmatic access to
the fit statistics calculation that allows one to modify its value during the
minimization.

For the case of $\chi^2$-statistics, the value to be minimized is
\begin{equation}
  \chi^2 = \sum_i^n (x_i - \mu_i)^2 / \sigma_i^2
\end{equation}
where the sum goes over all spectral channels, $i$, and where $x_i$
are the data counts with uncertainties, $\sigma_i$, and $\mu_i$ the
counts predicted by the model. We now modify $\chi^2$ to rise with
increasing difference of the accretion and X-ray luminosities,
\begin{equation} \label{chi}
  \chi^2_\mathrm{final} = \chi^2 \times
\left(1 + C\,\left(\dfrac{L_\mathrm{acc} - L_\mathrm{X} }{\Delta
  L_\mathrm{X} } \right)^{\gamma}\right).
\end{equation}
Here $\Delta L_\mathrm{X}$ is the uncertainty on the source luminosity
introduced by the uncertain distance to the object and other factors such as the
lack of knowledge of the emission pattern and accretion column geometry, while
$C$ and $\gamma$ are parameters adjusted in order to optimize the convergence
behavior of the $\chi^2$-minimization algorithm and to determine the strength of
the constraint. Our numerical investigations suggest to set $C{\sim}5$ and
$\gamma {\sim}4$, which leads to quick convergence when using a
Levenberg-Marquardt method. We found that these numbers worked well for the
datasets discussed in this paper. However, they are not guaranteed to be ideal
for every dataset, depending on, e.g., the used minimization algorithm and the
number of degrees of freedom, adjustments for quicker convergence might become
necessary. We settled for this implementation due to its simplicity and its
satisfying convergence during fitting (see also Fig.\,\ref{fig:con}).  
We emphasize that such an approach can be a powerful tool to include
additional and potentially complex constraints on the parameter landscape that
can not easily be included during model evaluation.

One possible disadvantage of this approach is that the more complex shape of the
$\chi^2$ landscape can lead to a tendency of the minimization algorithm to be
partly stuck in local minima. Also, $\chi^2_\mathrm{final}$ approximates $
\chi^2 $ only close to its minima, where energy is conserved. Consequently this
approach should only be used as a tool to find a set of physical best-fit
parameters and not to calculate statistical quantities, such as confidence
intervals. We validated the reliability of our new approach by reproducing a
previous successfully applicable of the model by BW07 to \her. As is described
in more detail in the appendix \ref{herx1}, we were able recover previous result
by \citet{wolff_16} with only minor deviations. \section{Application to
Cen~X-3}\label{res}

\begin{figure}
  \resizebox{\hsize}{!}{\includegraphics{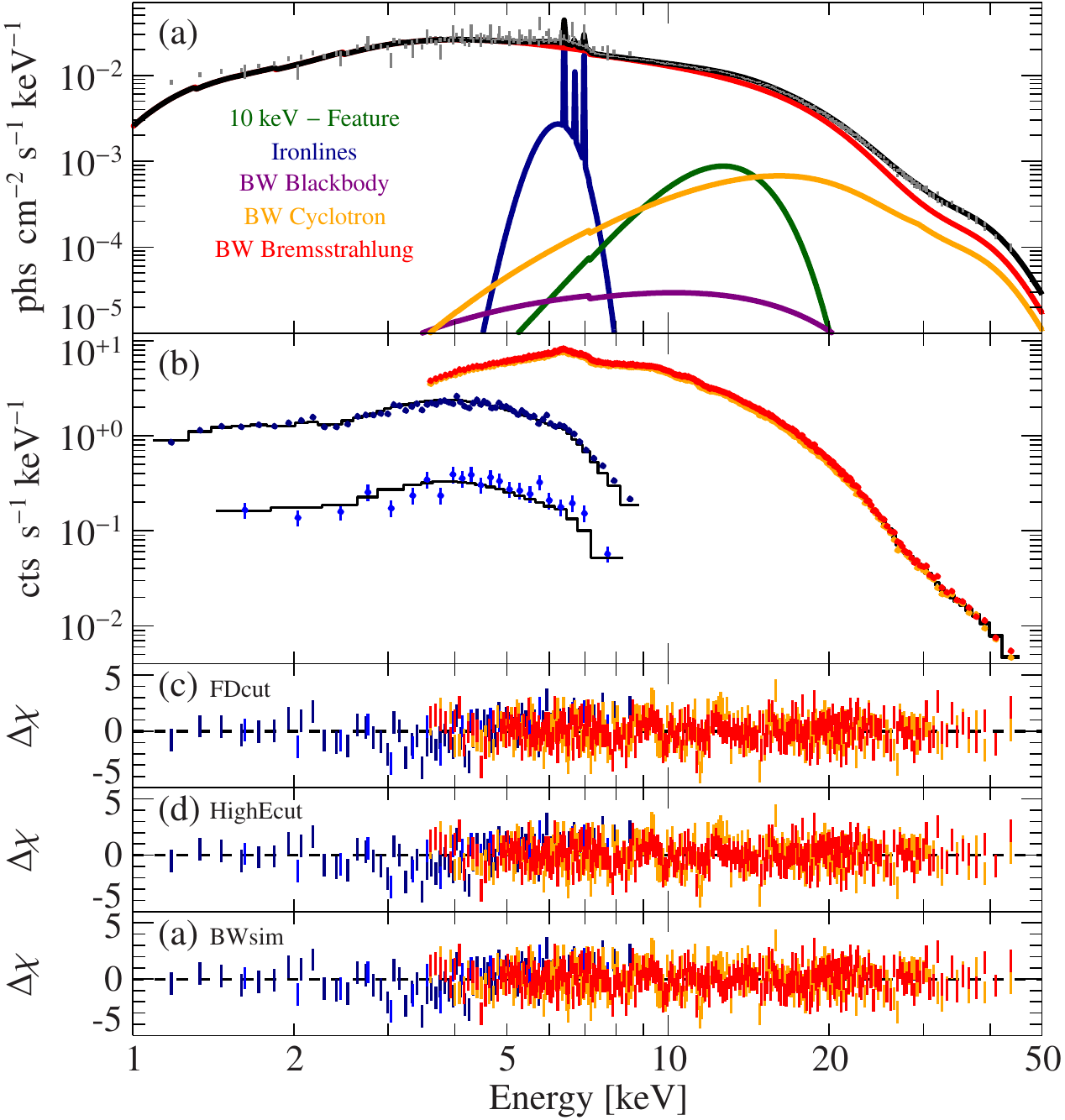}}
  \caption{Spectrum of \cen as seen by \nustar and \swift. \textit{(a)} Unfolded spectrum and the different components of
    the \texttt{BWsim} best-fit model. The difference between the
    unfolded data points and the model is caused by the algorithm used
    to unfold the data. In black the full spectrum is shown
    and colored are the different components. \textit{(b)} The second
    panel shows the spectrum fitted with \texttt{BWsim}. Red and
    orange indicate the data from the two focal plane modules of
    \nustar and shown in dark and light blue is the \swift data in
    window timing and photon counting mode. \textit{(c--e)}  Residuals for the \texttt{FDcut},
    \texttt{HighEcut} and the \texttt{BWsim} model.}\label{spec}
\end{figure}%

\begin{figure}
  \resizebox{\hsize}{!}{\includegraphics{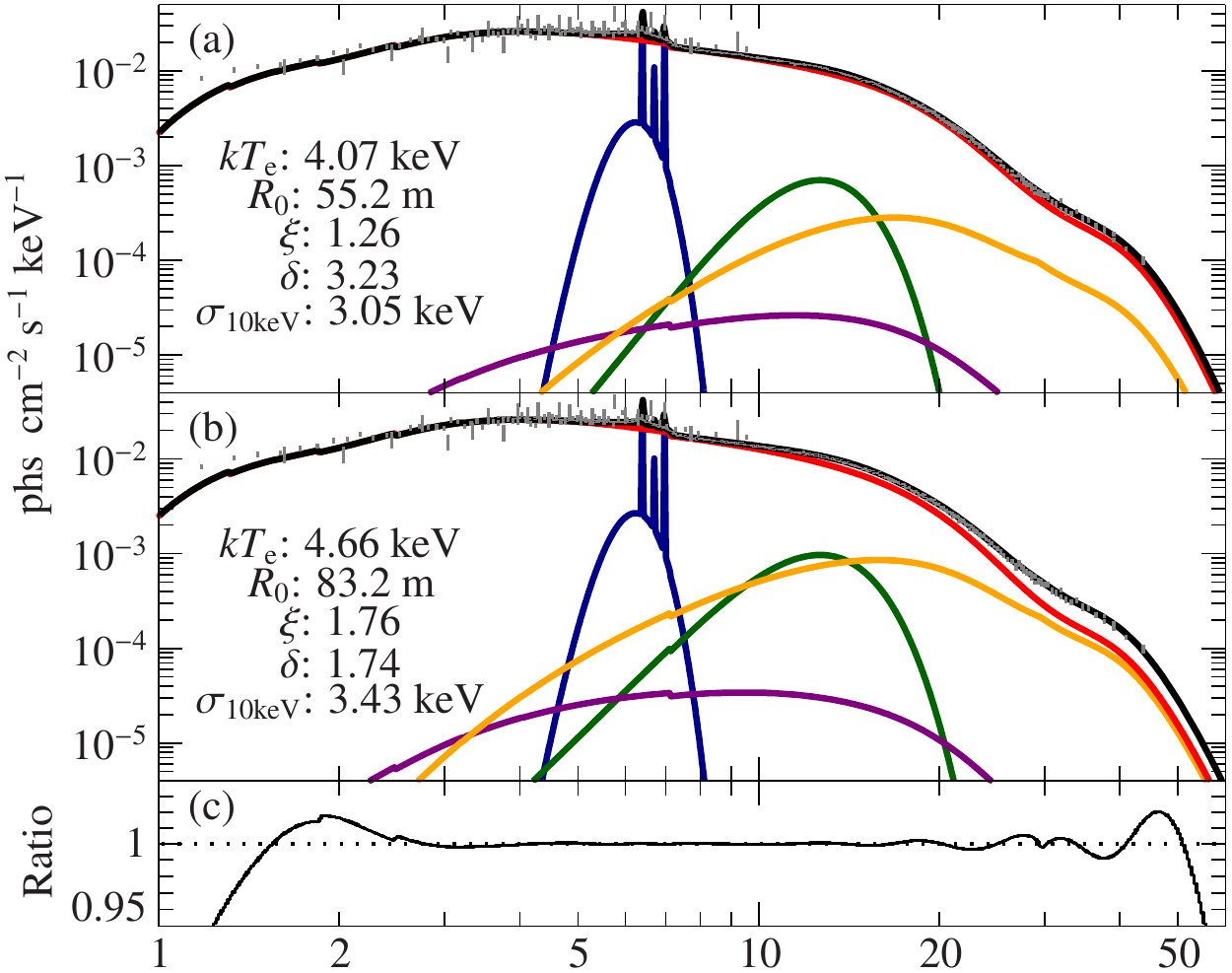}}
  \caption{ Similar to the upper panel in Fig.\,\ref{spec} the two plots of the
  unfolded spectrum display individual model components  and demonstrate the
  spectral change along the parameter correlation. \textit{(a)} 
  Model components for a small column radius. \textit{(b)} Model components for a
  larger column radius. \textit{(c)} The ratio between the high and low radius model.
  }\label{spec_comp}
\end{figure}

Having confirmed our novel approach, we continue to analyze the \cen
observation. In order to allow us to compare the spectral shape with
earlier observations, we first fitted the \textit{Swift}-XRT,
\nustar-FPMA, and -FPMB spectra with a
phenomenological model, i.e., a power-law with an exponential cut-off.
Previous studies \citep[e.g.,][]{Bissinger_2020,mueller_0115} have
shown the significant impact that implementation of the empirical
cut-off can have on the cyclotron line parameters, which can lead to
unphysically broad and deep lines. In order to probe such a bias, the
cut-off was modeled with the two commonly used models: First
\texttt{fdcut}, which is defined as
\begin{equation}
  F(E)\,\propto\, E^{-\Gamma} \frac{1}{1+\exp \left[{\left(E-E_{\mathrm{cut}}\right) / E_{\mathrm{fold}}}\right]}
\end{equation}
and second \texttt{powerlaw}$\times$\texttt{HighEcut}, defined as 
\begin{equation}
  \begin{aligned}
    F(E) \,\propto\, E^{-\Gamma}\begin{cases}
      \exp \left[\left(E_\mathrm{cut}-E\right) / E_\mathrm{fold}\right] & E \geq E_\mathrm{c} \\
     1.0 & E \leq E_\mathrm{cut}
    \end{cases} 
    \end{aligned}
\end{equation}
with the photon index $\Gamma$, the folding energy $E_\mathrm{fold}$
and a cutoff energy $E_\mathrm{cut}$. With a
$\chi^2_\mathrm{red}(\mathrm{d.o.f})$ of 1.32(450) and 1.28(453),
respectively, both models describe the spectrum reasonably well. Since
\texttt{HighEcut} has a discontinuity at the cut-off energy we smooth
this region with an additional absorption component tied to that
energy. Incidentally, this gets rid of some residuals caused by
calibration issues between 10\,keV and 14\,keV \citep[see][and
references therein]{wolff_16}, leading to a slightly lower
$\chi^2_\mathrm{red}$ value. As noted previously by, e.g.,
\cite{bruderi_2000}, \cen features very strong fluorescent iron lines
between 6 and 7\,keV. Most prominent is a neutral K$\alpha$ line at
6.4\,keV, but there are also lines from higher ionization states of
He- and H-like iron at 6.7\,keV and 6.97\,keV, which are also seen in
this observation. Additionally a broad iron line feature is necessary
to fully account for the iron line complex.
A similar feature is also present in the spectrum of \her. While
\citet{ebisame_asca_1996,Naik_2012} did not require such an feature, a similarly
broad iron line is seen in \her. \citet{asami_14} discussed several possible
origins for such a feature, namely unresolved emission lines, Comptonization in
an accretion disk corona and Doppler broadening at the inner disk or due to the
accretion stream. Similar consideration could be made for \cen but would exceed
the scope of this paper. 

The final model includes a calibration constant, $C_\mathrm{det}$, as
well as a partial
absorber\footnote{\url{https://pulsar.sternwarte.uni-erlangen.de/wilms/research/tbabs/}},
one broad and three narrow Gaussian iron
lines\footnote{\label{gaus}\url{https://heasarc.gsfc.nasa.gov/xanadu/xspec/manual/node177.html}}
that have a width smaller than the detector resolution,  a broad
Gaussian component\footnotemark[7], the so-called ``10 keV
feature''\footnotemark[7], and the cyclotron resonance feature around
30 keV, which we model by a multiplicative Gaussian absorption
component
(\texttt{gabs}\footnote{\url{https://heasarc.gsfc.nasa.gov/xanadu/xspec/manual/node240.html}}).
A similar, preliminary fit to \suzaku data was presented by \citet{marcu_2015} and will be published by Marcu-Cheatham et al.~(in prep).
In summary the full model is
\begin{multline}
        \mathtt{C}_\mathtt{det}\times \mathtt{tbnew}_\mathtt{feo}\times  \mathtt{tbnew}_ \mathtt{pcf} \times \\
        (    \mathtt{cont}+ \mathtt{gaussian}_\mathtt{3+1\, iron lines}+\\
        \mathtt{gaussian}_ \mathtt{10 keV}) \times  \mathtt{gabs},
\end{multline}
where $\texttt{cont}$ stands for the respective empirical continuum
model. The parameters values and 90\% uncertainties for the
corresponding best-fits are given in Table~\ref{para2}. The residuals
for these empirical fits are shown in Fig.~\ref{spec} (first two
residual panels). In a similar work, \citet{tomar_2020} recently
performed phenomenological fits to the same \nustar data of \cen as
used by us. To describe  the continuum they considered a physical
fit based on the theory of BW07. As their focus lay, however, on the
evolution of the CRSF and not a physical description of the continuum,
they choose a smooth high-energy cut-off model, \texttt{newhcut},
which also led to a lower reduced $\chi^2$. Due to the different
applied models, our ability to compare these results is limited.
However, \texttt{newhcut} is a smoothed version of \texttt{HighEcut}
and we would therefore expect our results not to deviate by too much.
Indeed, \citet{tomar_2020} derive a photon index of $1.21\pm0.01$,
close to the value of $1.17^{+0.10}_{-0.07}$ we found with our
\texttt{HighEcut} model. Overall, the most notable differences might
be the lack of a ``10-keV'' feature in the analysis by
\citet{tomar_2020}. We consider it likely that the reason for this is
the smoothed out transition region of the \texttt{newhcut} compared to
the hard break in the \texttt{HighEcut} model. Assuming a fixed
``smoothing width'' of 5 keV, \citet{tomar_2020} find a cut-off
energy, $E_\mathrm{cut}=14.14\pm0.06$\,keV. This result places the
transition region at a similar position and width as our ``10-keV''
feature. A more detailed comparison would be needed to
pin-down the precise differences. Further, the CRSF energy derived
from our analysis is consistent with the value of
$29.22^{+0.28}_{-0.27}$\,keV derived by \citet{tomar_2020}, further
indicating that the CRSF energy is well constrained by the \nustar
data.

As a next step, we apply the physical model {\texttt{BWsim}} to the
spectra. The full model is defined as
\begin{multline}\label{mod}
  \mathtt{C}_\mathtt{det}\times \mathtt{tbnew}_\mathtt{feo}\times \mathtt{tbnew}_\mathtt{pcf}\times\\
(\mathtt{zashift}((\mathtt{BWsim}+\mathtt{gaussian}_ \mathtt{10 keV})\times \mathtt{gabs})\\
+\mathtt{gaussian}_\mathtt{3+1\,iron lines}).
\end{multline}
Since \texttt{BWsim} calculates the emitted radiation in the reference
frame of the NS, it is necessary to account for the
gravitational redshift, which is done with the \texttt{zashift}
component. \texttt{BWsim} alone could not account for the ``10-keV''
feature and a broad Gaussian had to be included as part of the
continuum. Initially the iterative approach was used to fit the
described model. The resulting best fit had a reduced $\chi^2$ of 1.32
with 601 d.o.f.. After the iteration converged $\dot{M}$ was fixed to
$1.67\times10^{17}$ g s$^{-1}$. With the new fitting approach, i.e.,
including a $\chi^2$-penalty (see Eq.~\ref{chi}) almost identical
results were obtained (see Table~\ref{para} for the best-fit
parameters utilizing both methods).

\begin{figure*} \centering
    \resizebox{\hsize}{!}{\includegraphics[width = 9cm]{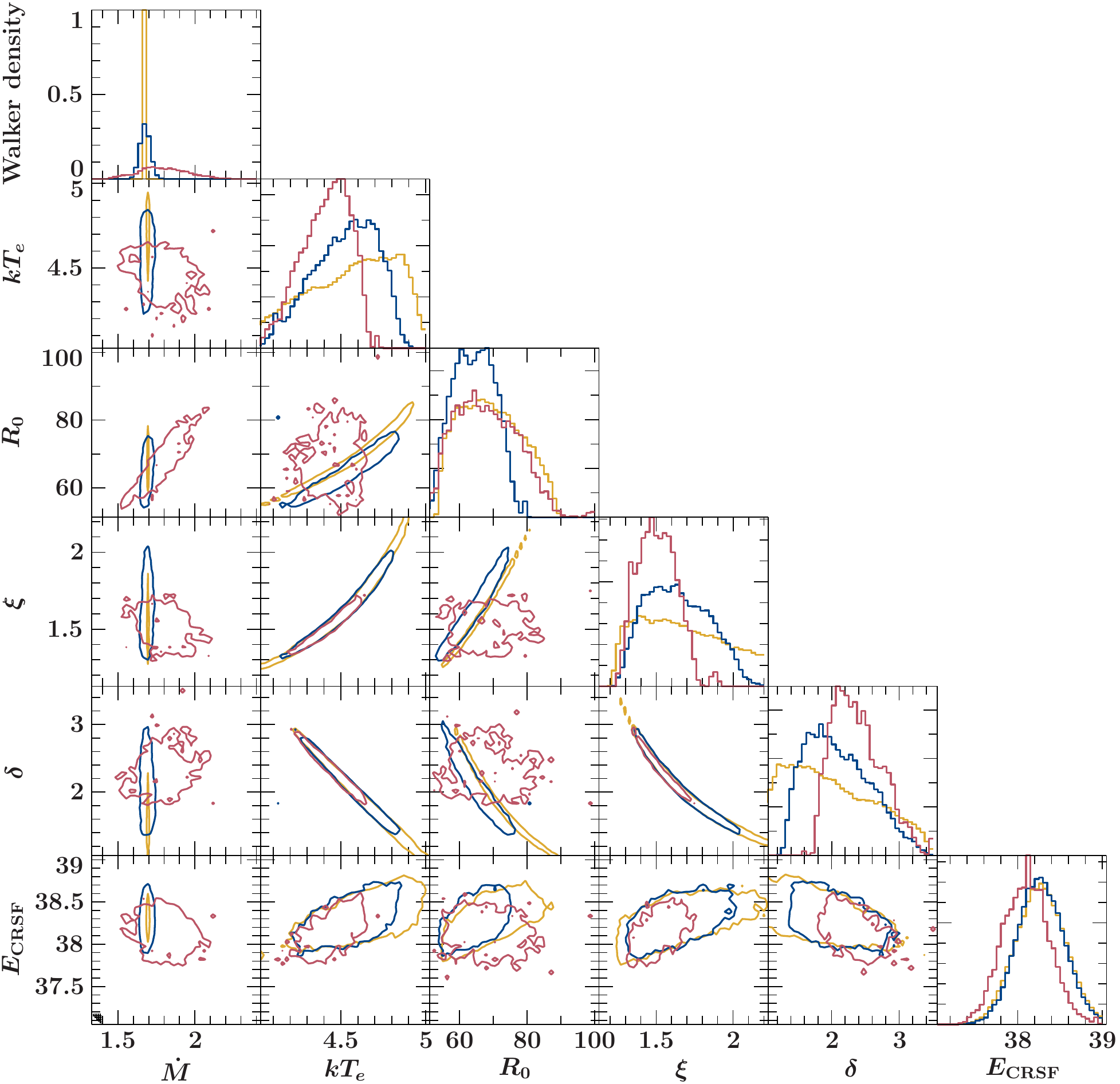}}
    \caption{ One- and two-dimensional marginal posterior density
      distributions derived from the MCMC walker distribution. In red
      the distributions for a free accretion rate, in blue the
      distribution under the influence of the constraint and in yellow
      with $\dot{M}$ fixed to the value found with the help of the
      constraint. The solid lines in the two-dimensional subplots
      denote the 1-sigma contours of the walker
      distribution.}\label{fig:mcmc_cons}
    \end{figure*}

To constrain parameter degeneracies, we therefore made use of a
Markov-chain Monte Carlo (MCMC) fitting approach, using the \texttt{ISIS}
implementation of the MCMC Hammer code \citep{mcmc_hammer_2013} by
M.A.\ Nowak. For this analysis each MCMC chain was run with ten walkers
per free parameter, ignoring the first 60\% of the chain to allow the
walker distribution to converge. In order to investigate the effect of
our constraint we performed three MCMC runs: Once with the accretion
rate as an once unconstrained free parameter, once with $\dot{M}$ fixed the
previously found best-fit value of
$1.67\times 10^{17}\,\mathrm{g}\,\mathrm{s}^{-1}$, and finally with
the new constraint applied. For the parameters of the BW07 model, the
resulting one- and two-dimensional marginal posterior density
distributions are shown in Fig.~\ref{fig:mcmc_cons}. It is apparent
that, without any constraint, only little information about the
accretion rate can be gained. Even though the distribution is still
well behaved, it is significantly wider than for the constrained fits.
Through a strong parameter degeneracy, this wider distribution
translates to a large uncertainty in $r_0$. The reason is that in the
unconstrained fits, $\dot{M}$ can shift away from values that conserve
energy, and therefore we also observe a systematic shift in most of
the other model parameters.

When applying the constraint, the distribution in $\dot{M}$ is
collapsed, even though further, smaller parameter degeneracies remain,
which are inherent to the BW07 model. When comparing the walker
distributions between our constraint and a fixed accretion rate, there
are only minor deviations.
 
To further illustrate these parameter correlations and the effect of
our constraint, we color-code a subset of the walkers by the
luminosities of their respective model. The result for the most
interesting parameters is shown in Fig.~\ref{fig:mcmc_lum}. The
continuum parameter not only clearly correlate with each other, but
also with the model luminosity, so that, e.g., a larger column radius
also leads to a higher luminosity. Further emphasized by the color
coding is the influence of the continuum parameter on the 10\,keV
feature and the CRSF, as both show a clear color gradient marking high
luminosities and consequently, e.g., high values of $r_0$. This
positive correlation between $r_0$ and the model luminosity might be
surprising, as an increase in $r_0$, with otherwise constant
parameters, reduces the luminosity of the BW07 model. However, to order
to still describe the data other continuum parameters have to change
as well, more than compensating the luminosity decrease that would
result from an increase in $r_0$ alone.

The resulting ``best-fit'' parameters of the MCMC analysis are shown
in Table~\ref{mcmc_table} as the median value and the 90\% confidence
intervals derived as 90\% quantiles of the run with a fixed accretion
rate. Figure~\ref{spec_comp} demonstrates how exactly the model
changes along the parameter degeneracy. With increasing $r_0$ the
Comptonized bremsstrahlung component in the model decreases in
dominance, while the thermal and cyclotron components, as well as the
10-keV feature, increase in strength. These changes compensate for the
flux lost in the decreased Comptonization component. These changes in
the continuum also drive the correlation between the strength and
width of the 10-keV feature, as these adjustments are required for a
satisfactory fit, since the feature is used to ``fill in'' the lost
flux in the 10\,keV band. Similarly, by investigating the correlation
between CRSF width and the BW parameters we see how the modeled
cyclotron line changes in width with the continuum, while its centroid
energy is comparably well constrained. As expected and seen in the
ratio plot, the spectral fluxes differ primarily at high or low
energies, where data are sparse or unavailable. These changes in
luminosity at the edges of the available energy range may also lead to
the discussed violation of energy conservation when $\dot{M}$ is just
fixed during $\chi^2$ minimization.

\begin{table}
  \caption{Results of the MCMC analysis. As best estimate we used the
    median value of each parameter chain and the 90\% percentile
    bounds from the projected 1D distributions.}
  \begin{tabular}{l r}\label{mcmc_table}
    Parameter &  MCMC best estimate  \\
    \hline
   $N_\mathrm{H1} \,[10^{22}\,\mathrm{cm}^{-2}]$ &  $1.45\pm0.07$\\
   $N_\mathrm{H2} \,[10^{22}\,\mathrm{cm}^{-2}] $&  $10.44^{+0.24}_{-0.23}$\\
   pcf                                &  $0.772\pm0.009$   \\
   $\dot{M}$ [10$^{17}$ g s$^{-1}$]  &  $^\dagger\,1.776$\\
   $kT_\mathrm{e} $ [keV]             &  $4.77^{+0.08}_{-0.09}$\\
   $r_0 $ [m]                         &  $77^{+5}_{-4}$ \\
   B ($10^{12} \mathrm{G}$)           &   $^\mathsection\,3.32$\\
   $\xi$                              &  $1.96^{+0.18}_{-0.16}$\\
   $\delta$                           &   $1.48^{+0.22}_{-0.20}$ \\
   A$_{6.4 \,\mathrm{ keV}}$ [$10^{-4}$ phs/cm$^2$/s]    &  $12.6\pm0.5$\\
   A$_{6.7\, \mathrm{ keV}}$ [$10^{-4}$ phs/cm$^2$/s]   &  $2.9\pm0.4$      \\
   A$_{6.9\, \mathrm{ keV}}$ [$10^{-4}$ phs/cm$^2$/s]    &  $6.0\pm0.4$     \\
   A$_{\mathrm{broad}}$ [ $10^{-4}$phs/cm$^2$/s]         &  $3.95\pm0.14$   \\
   E$_{\mathrm{broad}}$ [keV]         &  $6.208\pm0.020$    \\
   $\sigma_{\mathrm{broad}}$ [keV]    &  $0.520\pm0.014$    \\
   $A_{10\, \mathrm{keV}}$  [phs/cm$^2$/s]            &  $0.0108^{+0.0013}_{-0.0011}$ \\
   $E_{10\, \mathrm{keV}}$  [keV]           &  $16.44^{+0.11}_{-0.12}$ \\
   $\sigma_{10\, \mathrm{keV}}$  [keV]      &  $3.65\pm0.15$ \\
   $E_{\mathrm{CRSF}}$ [keV]          &  $^\ddag 38.37\pm0.12$ \\
   $\sigma_{\mathrm{CRSF}}$ [keV]     &  $8.38^{+0.15}_{-0.14}$ \\
   D$_{\mathrm{CRSF}}$          [keV]        &  $20.7^{+0.8}_{-0.7}$\\
   $C_{\mathrm{FPMB}}$                &  $^\dagger\,1.0255$   \\
   $C_{\mathrm{PC}}$                  &  $0.759\pm0.014$     \\
   $C_{\mathrm{WT}}$                  &  $^\dagger\,1.075$   \\
   \hline
   \tiny Note: $^\dagger$ fixed; \\
   \tiny $^\ddag$ In the NS rest frame;\\
   \,\,\tiny 29.46 keV for an external observer\\
   \tiny$^\mathsection$derived from $ B_{12} \simeq E_{\mathrm{CRSF}} / 11.6\, \mathrm{keV}$\\

  \(\)
  \end{tabular}
  \end{table}

\section{Discussion}\label{disc}

In this paper we have applied several different modeling approaches to
the \nustar and \swift spectra of \cen. Our phenomenological approach
confirms the stable spectral parameters of \cen over the last decade
\citep[][and references therein]{suchy_08,Naik_2012}.

In order to apply \texttt{BWsim} in any meaningful way, one has to
constrain its parameter $\dot{M}$. To this end we require
$L_\mathrm{X}$ and $L_\mathrm{acc}$ to be equal, which, however, is
only the case if the released energy is emitted isotropically. This
has been shown to be generally not the case \citep[see, e.g.,
Fig.~3.16 of][]{falkner_18}. Unfortunately, the value obtained with
this constraint is still the best guess, as the information on the
geometry of the system, which is necessary to improve this assumption,
such as the inclination of the NS's rotational axis or the
emissivity pattern of the accretion columns, is lacking.

In the model by BW07, the geometry of the accretion column is fully
described by the column radius, $r_0$. Interestingly our best-fit for
\cen shows a column radius of just $63\,\mathrm{m}$. This is just a
fraction of the $730\,\mathrm{m}$ estimated by BW07 in a ``fit by
eye'' of the model to earlier data of \cen. While these authors used
earlier observations with a ${\sim}3.5$ times higher unabsorbed
luminosity \citep{bruderi_2000}, the varying luminosity can only
account for $\sim$20\% of the variation of the column radius according
to Eq.~112 in BW07.
    
Another application of BW07 to \cen has been presented by
\citet{gottlieb_poster_2016} and will be published by Marcu-Cheatham et al.~(in prep).
These fits yield a column radius of around 65\,m,
in line with the results presented here. The large deviation between
the column radius presented here and that found by BW07 may partially
originate from the larger distance to \cen of 8\, kpc used in previous
studies, while we used the newer estimate of 5.7\, kpc by
\citet{distance57}. Consequently this leads to a higher inferred
luminosity and accretion rate, and, as shown in the parameter
correlations, a higher $\dot{M}$ will require a larger $r_0$ to model
the same spectrum. As also mentioned previously, during that
observation the source had about twice the 1--80\,keV unabsorbed
luminosity ($4.0 \times 10^{37}\,\mathrm{erg}\,\mathrm{s}^{-1}$) than
in our observation
($2.1 \times 10^{37}\,\mathrm{erg}\,\mathrm{s}^{-1}$). One has to be
careful when comparing observations at different unabsorbed
luminosities as the clumpy accretion stream does lead to varying
accretion rates which affect the structure of the accretion column.

Following \cite{davidson73}, assuming a dipole geometry one can
estimate the column radius from the accretion rate and the magnetic
field strength,
\begin{equation}
  r_{0} \sim R_{*}^{3/2} R_\mathrm{m}^{-1/2},
\end{equation} 
where $R_{*}$ is the radius of the NS and $R_\mathrm{m}$ is
radius where the plasma couples to the magnetic field lines. This
radius can be expressed in terms of the classical Alfv\'{e}n radius
$R_\mathrm{a}$ as $R_\mathrm{m} = k R_\mathrm{a}$ with
\begin{multline}R_\mathrm{a} = (2.4 \times 10^8 \mathrm{cm})\left(
\dfrac{B_0}{10^{12} \mathrm{G}} \right)^{4/7} \left( \dfrac{R_{*}}{10
\mathrm{\,km}} \right)^{12/7}\\ 
\times \left( \dfrac{\dot{M}}{10^{17}
\mathrm{\,g\, s}^{-1}} \right)^{-2/7}\left(
\dfrac{M_{*}}{M_{\odot}} \right)^{-1/7},
\label{eq:r0} \end{multline} where $B_0$ is the magnetic field
strength, $\dot{M}$ is the accretion rate, and $M_{*}$ is the mass of the
NS. Estimates for the parameter $k$ typically vary between 0.5--1
\citep{ghosh_79,wang_96,romanova_08}. Using this range and the parameters of
Cen~X-3 yields $R_\mathrm{m}$ of 2000--4000\,km and $r_0$ of 
$500$--$700\,\mathrm{m}$, far wider than our best-fit would suggest, but in good 
agreement with the results obtained by BW07 for this source.

There are many assumptions that go into the theory of BW07 that could
significantly influence the estimated column geometry, specifically
its radius. One assumption would be the presence of a more complex
accretion structure. For misaligned magnetic and accretion disk axes
one would expect a hollow accretion column or an accretion curtain
\citep{basko_1976,west_2017,Campbell_2012,Perna_2006}. Such a geometry
would lead to a much thinner accretion stream than the solid column
that is assumed in the model of BW07. It is reasonable to believe that
such a thin, elongated stream would produce a spectrum that is best
approximated by a thin accretion column within \texttt{BWsim}, as both
geometries would have a similar effects on parameters such as the
escape time of photons inside the column. A very similar model of a
hollow accretion column was applied to \cen by \citet{west_2017},
who derived an outer column radius of $\sim$750\,m, in much better
agreement with the value derived from Eq.~\ref{eq:r0}, and a width for
the column wall of $\sim$100\,m, which is in the order of our column
radius.

Another mechanism that can have a significant effect on the spectrum
and is not yet included in the BW07 is the interception of radiation
from the accretion column by the NS surface and its
subsequent reprocessing in the NS atmosphere. This
``reflection'' model was proposed by \citet{poutanen_13} as a possible
explanation for the origination of cyclotron lines in spectra of
accreting X-ray pulsars. \citet{lutovinov_15} have successfully
applied the reflection model to the observational data of the X-ray
transient V 0332+53 to describe the variability of the cyclotron line
parameters. \citet{postnov_15} show that reflection from the neutron
star surface can also lead to significant hardening of the spectrum in
the case of a filled accretion column. In order to estimate the
importance of this effect, one needs to know the spectrum of radiation
emitted at different heights of the column, as well as the angular
distribution of the emission. This information is currently not
included in the BW07 model, and therefore modeling the reflection from
the NS surface is outside the scope of this work. We note
that in the case of a hollow accretion column, the stopping and main
emission occur much closer to the surface and the reflection has no
noticeable effect on the continuum \citep{postnov_15}.

Finally, we address the validity of some of the fitted CRSF
parameters. In agreement with past measurements \citep[see, e.g.,
Fig.~5 of][for an overview of past CRSF measurements of
\cen]{tomar_2020}, the CRSF was found in all fits at $\sim$30\,keV and
with a width varying between 3.8 and 7.9\,keV.
The expected width of the cyclotron line can be estimated by
calculating the Doppler shift due to thermal broadening
\citep{schwarm_17,meszaros_85},
\begin{equation}
 \frac{\Delta E}{E}=\sqrt{8 \ln 2 \frac{k_{\mathrm{B}} T}{m_{\mathrm{e}} c^{2}}} \cos \theta
\end{equation}
where $E$ is the cyclotron energy, $\theta$ is the viewing angle
relative to the magnetic moment, and $T$ is the electron temperature.
Using the electron temperature derived from the physical fits, we find
$\Delta E \sim 6.5\,\mathrm{keV} |\cos \theta| $. Although the average
viewing angle is hard to constrain, this estimate indicates that, at
least for our physical fits, the CRSF seems to be slightly wider than
expected from thermal broadening, This could again be a result of the
choice of continuum, as we have seen that this can significantly
influence the shape of the cyclotron line. Alternatively, such
additional broadening might also be induced by varying magnetic field
strength along the height of the column and/or by a varying
temperature profile within the column \citep{west_2017,falkner_18}.
Taking all these assumptions into account, the electron temperature
falls well inside the expected range of values.

As a quick check of some of the model assumptions, we can calculate
the integration height $z_\mathrm{max}$ and the height of the sonic
point $z_\mathrm{sp}$ according to Eq.~80 and Eq.~31 of BW07, as most
of the emission is expected to occur below the sonic point. One finds
$z_\mathrm{max} = 5.5$\,km and $z_\mathrm{sp} = 2.8$\,km. Importantly
the sonic shock lies within the integration region, as it is supposed
to be. Further, following \citet{wolff_16}, the ``blooming fraction''
at the sonic point can be obtained,
$(( R_{*} + z_\mathrm{sp}) / R_{*})^3 = 2.1$, which clearly exceeds
unity, hinting at a deviation from a simple cylindrical geometry.

Similarly, assuming a dipole magnetic field, we can calculate the
magnetic field strength at the sonic point to be $1.6\,B_{12}$, which
also indicates that the assumption of a constant magnetic field might
introduce systematic errors. However, to which degree this actually
affects our results is hard to specify as the emission intensity along
the column height is not directly accessible.

Finally, from the similarity parameters $\xi$ and $\delta$, it is
possible to derive the average electron scattering cross-section
$\bar{\sigma}=2.55\times 10 ^{-4} \sigma_\mathrm{T}$ and the
cross-section parallel to the magnetic field
$\sigma_{\|} =3.12\times 10 ^{-5} \sigma_\mathrm{T}$, while the
cross-section perpendicular to the magnetic field is frozen to the
Thomson cross-section $\sigma_\mathrm{T}$. These are ordered as one
would expect $ \sigma_{\|}<\bar{\sigma}<\sigma_{\perp}$ and are
generally in the order of, but some-what lower than, the cross-section
derived by BW07.

In summary, while the BW07 model describes the physical processes of
the formation of the radiation well, its assumption of a filled
cylindrical accretion column of constant radius is likely to introduce
systematic errors in the geometry parameters when comparing the model
with data. Further development of the model toward a more realistic
column structure is therefore needed. We emphasize that despite these
simplifications, besides the implementation by \citet{Farinelli_2012},
the model is still the only physics-based model for accretion column
emission applicable to fitting observational data and that the major
conclusions drawn from fits of the BW07 model, e.g., the dominance of
bulk motion Comptonization or bremsstrahlung emission on the observed
spectra, will be largely unaffected by the simplified assumptions for
the accretion geometry.

\section{Conclusions} \label{conc}

In this paper we presented a detailed spectral analysis of combined
\nustar and \swift spectra of \cen. To simplify application of the
physical column model by BW07 to the spectrum of an accreting pulsar,
a new approach of fitting was introduced. By biasing the value of
$\chi^2$ toward physically consistent parameters, it was possible to
apply the physical model derived by BW07 without having to assure
energy conservation in any other, external way. This approach allows one to explore non standard parameter spaces with efficient $\chi^2$-minimization algorithms alternatively to more computationally expensive sampling methods such as MCMC. In 
many cases,  this will considerably simplify the application of the BW07 on
observational data but this is also generally applicable to complex spectral models where certain parameter combinations may lead to the violation of model assumptions. With this new approach previous fits to the
spectrum of \her were reproduced and the spectrum of \cen could be
fitted with similar quality as conventional phenomenological models.
The familiar features which have been seen in previous studies were
found in our analysis as well, e.g., the prominent iron lines, the
CRSF at $\sim$30\, keV and a partially absorbed power-law continuum.
The parameters of \texttt{BWsim} describing the accretion column
suggest a thin or possibly a hollow accretion column or curtain.

\begin{acknowledgements}
  This work has been funded in part by the Bundesministerium f\"ur
  Wirtschaft und Technologie through Deutsches Zentrum f\"ur Luft- und
  Raumfahrt grants 50\,OR\,1909. This work used data from the \nustar
  mission, a project led by the California Institute of Technology,
  managed by the Jet Propulsion Laboratory, and funded by the National
  Aeronautics and Space Administration. The material is based upon
  work supported by NASA under award number 80GSFC21M0002. ESL
  acknowledges support by Deutsche Forschungsgemeinschaft grant
  WI1860/11-1 and RFBR grant 18-502-12025. This research also made
  use of the \nustar Data Analysis Software (NuSTARDAS) jointly
  developed by the ASI Science Data Center (ASDC, Italy) and the
  California Institute of Technology (USA), as well as data from the
  \swift satellite, a NASA mission managed by the Goddard Space Flight
  Center. MTW is supported by the NASA Astrophysics Explorers Program
  and the NuSTAR Guest Investigator Program. This research also has
  made use of a collection of ISIS functions (ISISscripts) provided by
  ECAP/Remeis observatory and MIT
  (http://www.sternwarte.uni-erlangen.de/isis/). Finally, we
  acknowledge the helpful advice concerning readability and the
  general support by Victoria Grinberg.
\end{acknowledgements}

\bibpunct{(}{)}{;}{a}{}{,} 
\bibliographystyle{aa} 
\bibliography{paper} 

\begin{appendix}

\def\arraystretch{1.5}
\section{Fit parameters}
\begin{table}[htb]
\caption{Best-fit parameters for empirical models applied to the \cen observation}
\begin{tabular}{lrr}\label{para2}

 Parameter & HighEcut& FDcut\\
 \hline
$N_\mathrm{H1} \,[10^{22}$cm$^{-2}]  $ & $1.71\pm0.28$      &  $1.62\pm0.28$ \\
$N_\mathrm{H2} \,[10^{22}$ cm$^{-2}]  $  & $10.8^{+1.1}_{-1.0}$       &  $10.3\pm1.0$ \\
$\mathrm{pcf}$ & $0.798^{+0.030}_{-0.033}$ &  $0.78\pm0.04$ \\
Norm$_\mathrm{PL}$  & $0.205^{+0.033}_{-0.022}$   &  $0.178^{+0.019}_{-0.015}$ \\
$\Gamma$  & $1.17^{+0.10}_{-0.07}$  &  $1.05\pm0.07$ \\
$E_{\mathrm{ cutoff}} $ [keV]  & $12.67^{+0.17}_{-0.15}$  &  $22^{+6}_{-5}$ \\
$E_{\mathrm{Fold}}$ [keV]  & $9.6^{+0.7}_{-0.4}$        &  $6.1^{+0.9}_{-1.3}$ \\
$\sigma_{\mathrm{gabs}}^\dagger$ [keV]  & $1.63^{+0.35}_{-0.27}$ &--- \\ 
D$_{\mathrm{gabs}}^\dagger$  [keV]  & $0.29^{+0.15}_{-0.10}$      & ---\\ 
A$_{6.4\,\mathrm{keV}}$   [$  10^{-4}$ phs/cm$^2$/s ]&$12.2^{+1.7}_{-1.8}$    &  $12.4^{+1.8}_{-1.9}$ \\
A$_{6.7\,\mathrm{keV}}$  [$  10^{-4}$ phs/cm$^2$/s] & $2.4\pm1.4$       &  $2.9^{+1.6}_{-1.5}$ \\
A$_{6.9\,\mathrm{keV}}$  [$  10^{-4}$ phs/cm$^2$/s] & $5.3\pm1.5$       & $5.9^{+1.6}_{-1.5}$ \\
A$_\mathrm{broad}$ [$  10^{-4}$]  & $43\pm6$       &  $39^{+6}_{-7}$ \\
E$_\mathrm{broad}  $ [keV] & $6.23\pm0.08$      &  $6.21^{+0.08}_{-0.09}$ \\
$\sigma_\mathrm{broad}$ [keV] & $0.55\pm0.06$      &  $0.51\pm0.06$ \\
A$_{10\,\mathrm{keV}}$   [$  10^{-4}$ phs/cm$^2$/s]  & $180^{+130}_{-70}$   &  $97^{+37}_{-27}$ \\
E$_{10\,\mathrm{keV}}$ [keV] & $13.5^{+1.0}_{-1.2}$       &  $12.6\pm0.4$ \\
$\sigma_{10\,\mathrm{keV}}$ [keV] & $4.1\pm0.6$        &  $2.9^{+0.5}_{-0.4}$ \\
E$_\mathrm{CRSF} $ [keV]  & $29.6^{+0.5}_{-0.4}$   &  $28.90^{+0.36}_{-0.29}$ \\
$\sigma_\mathrm{CRSF} $  [keV]& $3.8\pm0.5$    &  $6.3^{+0.9}_{-1.1}$ \\
D$_\mathrm{CRSF}$  [keV]   & $2.7\pm0.7$ &  $13^{+10}_{-6}$ \\
C$_\mathrm{FPMB}$  & $1.0256\pm0.0022$ &  $1.0255\pm0.0022$ \\
C$_\mathrm{PC}$  & $0.76\pm0.06$     &  $0.76\pm0.06$ \\
C$_\mathrm{WT}$  & $1.076\pm0.017$   &  $1.075\pm0.017$ \\
$\chi_{\mathrm{red}}^{2} / \mathrm{d.o.f.}$ &  1.24 / 596  & 1.31 / 598 \\
\hline
\end{tabular}
\tiny Note: $^\dagger$ energy fixed to cutoff energy
\end{table}

\begin{table}[htb]
\vspace{1cm}
\caption{Best-fit parameters for \texttt{BWsim} model applied to the \cen observation}
\begin{tabular}{lrr}\label{para}

  Parameter &  $\chi^2$ biased method& iterative method  \\
  \hline
 $N_\mathrm{H1} \,[10^{22}$cm$^{-2}]$              & $1.37^{+0.28}_{-0.26}$        & $1.36^{+0.29}_{-0.25}$ \\
 $N_\mathrm{H2}\,[10^{22}$cm$^{-2}]$          & $10.3^{+1.0}_{-0.7}$          &$10.3^{+1.1}_{-0.6}$\\
 pcf                                    & $0.77\pm0.04$     & $0.78\pm0.04$\\
 $\dot{M}$ [10$^{17}$ g s$^{-1}$ ]      & $1.67^{+0.05}_{-0.06}$     & $^\dagger\,1.67$\\
 $kT_\mathrm{e} $ [keV]                 & $4.53^{+0.11}_{-0.13}$        & $4.51^{+0.10}_{-0.08}$ \\
 $r_0 $ [m]                             & $63.4^{+2.7}_{-3.9}$                            & $67^{+6}_{-9}$  \\
 B ($10^{12} \mathrm{G}$)               &      $^\mathsection\,3.32$ &  $^\mathsection\,3.32$\\
 $\xi$                                  & $1.61^{+0.14}_{-0.12}$    & $1.58^{+0.13}_{-0.11}$\\
 $\delta$                               &$2.04^{+0.32}_{-0.23}$     & $2.10^{+0.23}_{-0.16}$ \\
 A$_{6.4 \,\mathrm{ keV}}$  [$  10^{-4}$]     & $12.5^{+1.9}_{-1.8}$    & $1s2.6^{+1.7}_{-1.9}$\\
 A$_{6.7\, \mathrm{ keV}}$   [$  10^{-4}$]    &$2.8^{+1.5}_{-1.4}$       &$2.9^{+1.5}_{-1.4}$\\
 A$_{6.9\, \mathrm{ keV}}$    [$  10^{-4}$]   &$6.0^{+1.6}_{-1.4}  $  & $6.1\pm1.5$\\
 A$_{\mathrm{broad}} $       [$  10^{-4}$]      & $39\pm6$     & $12.6^{+1.7}_{-1.9}$ \\
 E$_{\mathrm{broad}}$ [keV]             & $6.21^{+0.08}_{-0.09}$      & $6.20^{+0.08}_{-0.09}$ \\
 $\sigma_{\mathrm{broad}}$ [keV]        & $0.52\pm0.06$       & $0.52^{+0.05}_{-0.06}$ \\
 $A_{10\, \mathrm{keV}}$    [$  10^{-4}$]   & $81^{+42}_{-22}$  & $79^{+43}_{-22}$ \\
 $E_{10\, \mathrm{keV}}$  [keV]         & $16.6^{+0.4}_{-0.5}$      & $16.6\pm0.4$\\
 $\sigma_{10\, \mathrm{keV}}$  [keV]    & $3.3^{+0.6}_{-0.5}$        & $3.3^{+0.6}_{-0.5}$ \\
 $E_{\mathrm{CRSF}}$ [keV]              & $^\ddag 38.15^{+0.43}_{-0.29}$      & $^\ddag 38.14^{+0.43}_{-0.27}$\\
 $\sigma_{\mathrm{CRSF}}$ [keV]         & $8.7^{+0.5}_{-0.4}$        & $8.8\pm0.4$ \\
 D$_{\mathrm{CRSF}}$  [keV]    &$22.3^{+2.1}_{-1.7}$  & $22.6^{+2.1}_{-1.8}$\\
 $C_{\mathrm{FPMB}}$    & $1.0255\pm0.0021$     & $1.0256\pm0.0022$ \\
 $C_{\mathrm{PC}}$      & $0.76\pm0.06$         & $0.76\pm0.06$\\
 $C_{\mathrm{WT}}$      & $1.075^{+0.016}_{-0.017}$     & $1.075\pm0.017$\\
 $\chi_{\mathrm{red}}^{2} / \mathrm{d.o.f.}$ &   1.32 / 601 &  1.32 / 601\\
 \hline
\end{tabular}\\
 \tiny Note: $^\dagger$ fixed; \\
 \tiny $^\ddag$ In the NS rest frame \\
  \,\,\tiny For an external observer 29.35 keV \& 29.34 keV respectively \\
 $^\mathsection$ tied to E$_{\mathrm{CRSF}}$

\end{table}

\begin{figure*} \centering
  \resizebox{\hsize}{!}{\includegraphics{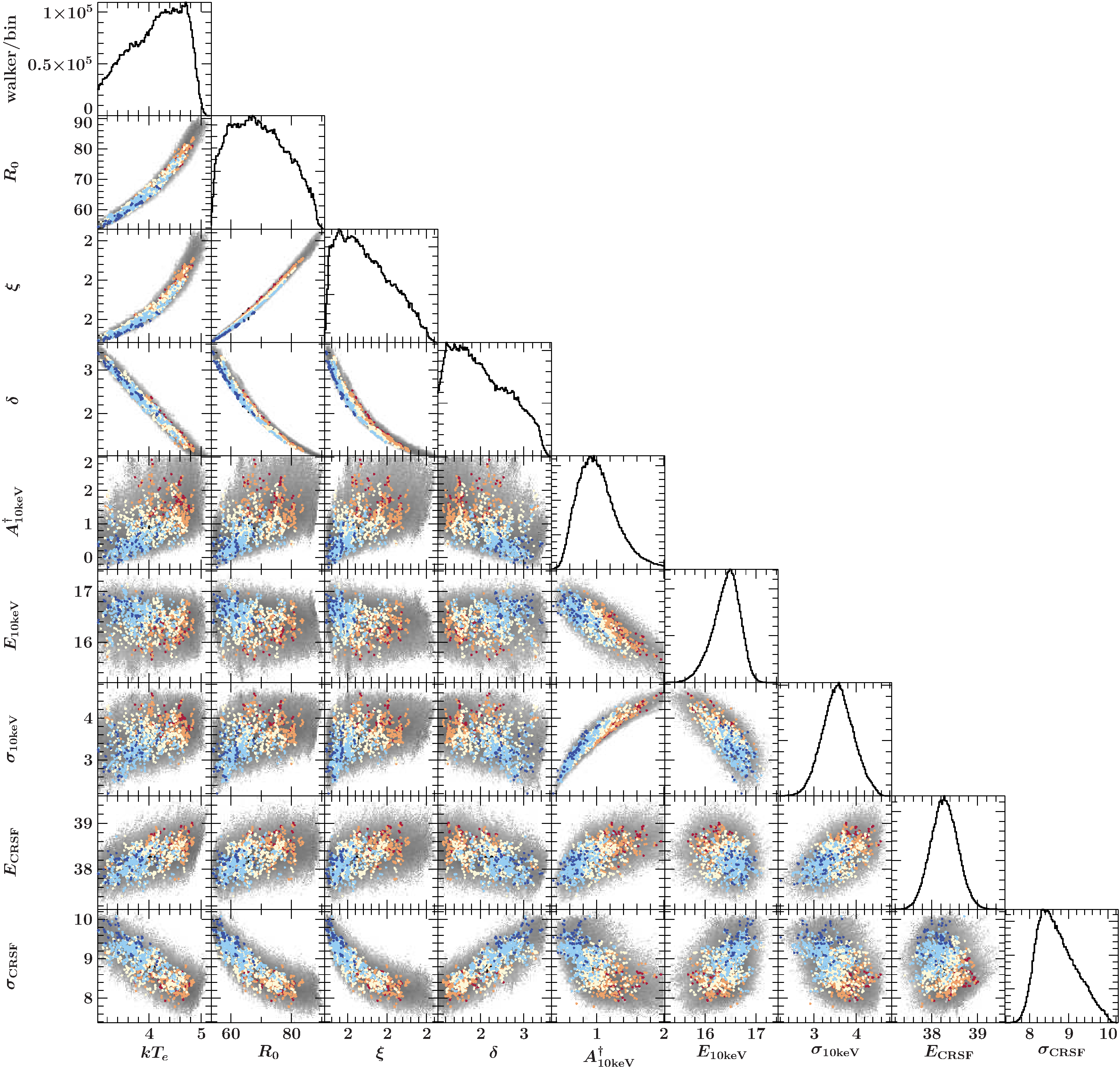}}
  \caption{ One- and two-dimensional marginal posterior density distributions
  derived from the MCMC walker distribution. For this run $\dot{M}$ was fixed. The under-laying gray contour shows the walker distribution of the converged chain. A random subset of walkers was picked out and color-coded by luminosity, with walkers corresponding to a bright model colored in red and less bright Walkers colored in blue.   
   $^\dagger$: Multiplied by 100. }\label{fig:mcmc_lum}
  \end{figure*}

\clearpage
  \section{Proof of concept: \her} \label{herx1}
  To validate our new approach, we applied it to archival data
  of Her~X-1 which was already successfully modeled by \cite{wolff_16} with
  their iterative approach, before then turning to the analysis of
  \textit{NuSTAR} and \textit{Swift} data of \cen.
  
  \begin{figure}
    \resizebox{\hsize}{!}{\includegraphics{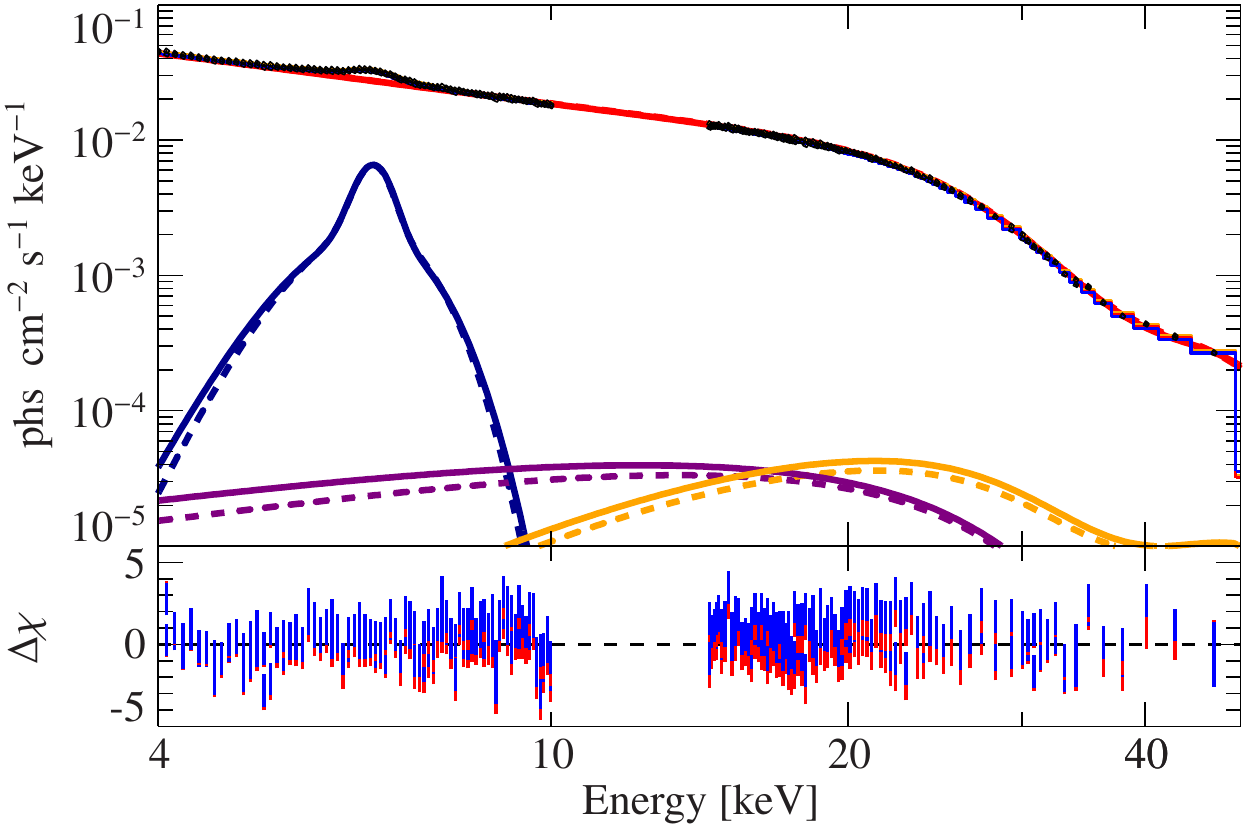}}
    \caption{Proof of concept: Comparison between the best fit to the spectrum of \her,
      obtained with the new fitting approach (solid lines in the upper panel
      and red residuals in the bottom panel) and the fit performed by
      \citet{wolff_16} (dashed lines in the upper panel and blue residuals in
      the bottom panel).The red component corresponds to Comptonized bremsstrahlung, the 
      orange one to Comptonized cyclotron emission, the violet one to the
      Comptonized blackbody emission, and the blue one 
      to the iron line complex.  The corresponding parameter values are
      shown in Tab. \ref{para_her}. }\label{fig:wolffcomp}
  \end{figure}
  
  \her is an intermediate-mass X-ray pulsar and the first source in
  which a cyclotron line was observed
  \citep{truemoer_her_1978}. It was observed by \nustar
  (ObsID 30002006005) on 2012 September 22, when it reached a luminosity
  of $\sim\,4.9{\times}\,10^{37}\,\mathrm{erg}\,\mathrm{s}^{-1}$,
  putting \her above  or close to the super-critical luminosity according to \citep{becker_2012}, As \citet{wolff_16} have estimated  a critical luminosity for \her of $L_{\text {crit }} \sim 7.3 \times 10^{36} \operatorname{erg} \mathrm{s}^{-1}$.
  
  \citet{wolff_16} excluded the region between 10\,keV and 14.4\,keV due
  to uncertainties in the calibration. For consistency, the same was done in this
  analysis. The comparison between the best fit, found with the new
  approach and the classical approach, shown in Fig.~\ref{fig:wolffcomp}
  and Table~\ref{para_her}, illustrates their convergence to
  equivalent fits. Any remaining deviations are consistent with being due to the
  slightly different calibration used for the data extraction and the
  fact that \citet{wolff_16} allowed the accretion rate to deviate by
  30\% from complete energy conservation.

  \begin{table}
    \caption{Proof of concept of our novel fitting method: Comparison of our
      best-fit \texttt{BWsim} parameters for \her and the ones found by
      \citet{wolff_16}. As discussed in Sect.~\ref{meth}, for the new approach the
      ``uncertainties'' have lost their strict meaning of confidence interval.
      They illustrate, however, that close to the chi-square minimum the
      distortion of the confidence intervals is small. The plot of the two models
      and the corresponding residuals are shown in Fig.~\ref{fig:wolffcomp}.
      }
    \begin{tabular}{lrr}\label{para_her}
   
      Parameter &  New approach &   \citet{wolff_16} \\
      \hline
     $N_\mathrm{H} \,[10^{22} $cm$^{-2} ]$         & $^\dagger\,1.7 $   & $^\dagger\,1.7 $\\
     $\dot{M}$ [10$^{17}$ g s$^{-1}$ ]  & $2.29^{+0.07}_{-0.04}$                  &$ ^\dagger2.5935 $\\
     $kT_\mathrm{e} $ [keV]             & $4.30^{+0.13}_{-0.10}$                  &$ 4.58_{-0.07}^{+0.07}$\\
     $r_0 $ [m]                         & $83^{+7}_{-5}$                          & $107.0_{-1.8}^{+1.7}$ \\
     B [$10^{12} \mathrm{G}$]  &       $^\mathsection\,4.16$ & $^\mathsection\, 4.25$ \\
     $\xi$                              & $1.295^{+0.028}_{-0.025}$               & $^\ddagger\, 1.355$  \\
     $\delta$                           & $2.37^{+0.12}_{-0.13}$                  & $^\ddagger\, 2.38$ \\
     $E_{\mathrm{CRSF}}$ [keV]          & $37.0^{+0.5}_{-0.4}$                    & $37.7_{-0.2}^{+0.2}$ \\
     $\sigma_{\mathrm{CRSF}}$ [keV]     & $6.7\pm0.4$                             & $7.1_{-0.2}^{+0.2}$ \\
     $D_{\mathrm{CRSF}}$   [keV]          & $0.89^{+0.14}_{-0.09}$                  & $0.98_{-0.06}^{+0.06}$\\
     A$_{\mathrm{FeK} \alpha_{\mathrm{b}}}$ [phs/cm$^2$/s]              & $0.0029_{-0.0006}^{+0.0006}$  & $0.0029_{-0.0006}^{+0.0006}$ \\
     E$_{\mathrm{FeK} \alpha_{\mathrm{b}}}$ [keV]         & $6.541^{+0.016}_{-0.017}$               & $6 .61_{-0.02}^{+0.02}$ \\
     $\sigma_{ (\mathrm{FeK} \alpha_{\mathrm{b}}}$ [keV]    & $0.25\pm0.04$                           & $0.26_{-0.04}^{+0.03}$ \\
     A$_{(\mathrm{FeK} \alpha_{\mathrm{b}}}$  [phs/cm$^2$/s]               & $0.0045\pm0.0006$                       & $0.0043_{-0.0055}^{+0.0005}$ \\
     E$_{\mathrm{FeK} \alpha_{\mathrm{b}}} $ [keV]         & $6.47^{+0.07}_{-0.10}$                  & $6.53_{-0.08}^{+0.07}$ \\
     $\sigma _{\mathrm{FeK} \alpha_{\mathrm{b}}}$ [keV]    & $0.92^{+0.20}_{-0.14}$                  & $0.90_{-0.14}^{+0.21}$ \\
     $\chi_{\mathrm{red}}^{2} / \mathrm{d.o.f.}$ & 1.20 / 310  & 1.63 /310 \\
     \hline
      \end{tabular}
      \tiny Note: $^\dagger$ fixed;   $^\ddagger$ derived;  $^\mathsection$ derived from E$_{\mathrm{CRSF}}$ by $ B_{12} \simeq E_{\mathrm{cyc}} / 11.6\, \mathrm{keV}$
    \end{table}

\end{appendix} 
 
\end{document}